\documentclass[showpacs,aps,graphicx,onecolumn]{revtex4}
\usepackage{amsmath}
\usepackage{mathrsfs}
\usepackage{amsfonts}
\usepackage{amssymb}
\usepackage{graphicx}
\usepackage{subfigure}
\usepackage{eufrak}
\usepackage{multirow}
\usepackage{longtable}
\usepackage{float}
\usepackage{bm}
\usepackage{xcolor}
\usepackage[colorlinks,linkcolor=blue,anchorcolor=blue,citecolor=blue,urlcolor=blue]{hyperref}
\usepackage{soul,color,xcolor}

\begin{document}

\title{Heralded and high-efficient entanglement concentrations based on linear optics assisted by time-delay degree of freedom}

\author{Gui-Long Jiang,\textsuperscript{1} Wen-Qiang Liu,\textsuperscript{2} and Hai-Rui Wei\textsuperscript{1,}}
\email[]{hrwei@ustb.edu.cn}
\address{\textsuperscript{\rm1} School of Mathematics and Physics, University of Science and Technology Beijing, Beijing 100083, China\\
\textsuperscript{\rm2} Center for Quantum Technology Research and Key Laboratory of Advanced Optoelectronic Quantum
Architecture and Measurements (MOE), School of Physics, Beijing Institute of Technology, Beijing 100081, China}

\begin{abstract}
Entanglement concentration is a critical technique to prevent degraded fidelity and security in long-distance quantum communication. We propose novel practical entanglement concentration protocols (ECPs) for less-entangled Bell and Greenberger-Horne-Zeilinger states with unknown parameters by solely using simple linear optics. We avoid the need for the post-selection principles or photon-number-resolving detector to identify the parity-check measurement completely by orchestrating auxiliary time degree of freedom, and the success of ECPs is exactly heralded by the detection signatures without destroying the incident qubits. Additionally, the outting incident photons kept are in the maximally entangled or the less-entangled state, and the success probability can be increased by recycling the latter. The heralded and the basic linear optical elements make our practical ECPs are accessible to experimental investigation with current technology.
\end{abstract}

\maketitle

\section{Introduction}\label{sec1}

Entanglement, as a unique quantum mechanical phenomenon, plays an essential role in quantum information processing (QIP) theory, and has attracted widespread attentions \cite{1,2}.
The entangled photons are often regarded as the most promising resource in a wide-range of long-distance quantum communication tasks:
quantum key distribution \cite{3,4}, quantum teleportation \cite{5,6,7}, quantum dense coding \cite{8,9}, quantum secret sharing \cite{10,11}, quantum secure direct communication \cite{12,13,14}, etc.
Maximally optical entangled states are indispensable for  most QIP applications, owning to their high-speed transmission and outstanding low-noise properties \cite{1}.
However, the maximally entangled states may be degraded to less-entangled states as the optical systems will inevitably interact with channel noise and their environment in an actual long-distance quantum communication, and these influences may degrade the fidelity, security, and success of the protocols.
Fortunately, such degraded entangled state problems can be remedied well by employing entanglement purification \cite{15} and entanglement concentration techniques \cite{16}.

Entanglement purification protocol (EPP) was first proposed by Bennett \emph{et al}. \cite{15} in 1996 with controlled-NOT gate to extract a two-photon pure singlet state in a Werner state.
Consequently, increasing efforts are being devoted to improving EPP \cite{17,18,19,20,21,22,23}.
Entanglement concentration protocol (ECP) \cite{16}  is another way to distill a maximally entangled state from a pure  less-entangled state. The first ECP, based on Schmidt decomposition, was proposed by Bennett \emph{et al}. \cite{16} in 1996.
Later in 1999, Bose \emph{et al}. \cite{Bose} proposed an ECP via entanglement swapping, 
and the improved work was proposed by Shi \emph{et al.} \cite{Shi} in 2000.
In 2001, Yamamoto \emph{et al}. \cite{Yamamoto} and Zhao \emph{et al}. \cite{Zhao} proposed an ECP for two partially entangled photon pairs with linear optics,
and the schemes were experimentally demonstrated later in 2003 \cite{experiment1,experiment2}.
In 2002, Paunkovoi\'{c} \emph{et al.} \cite{Paunkov} proposed an ECP based on quantum statistics, and less knowledge of the initial states is required than  most linear-optics-based ECPs.
In 2008 and 2012, Sheng \emph{et al}. \cite{Sheng2008,Sheng2012} proposed ECPs by exploiting cross-Kerr medium. The efficiency of such ECPs is higher than the linear-optics-based ones.

The existing ECPs are mainly focused on two-photon Bell states, and they are also suitable to multi-photon Greenberger-Horne-Zeilinger  (GHZ) states \cite{Sheng2008,Sheng2012}, but not compatible with $W$ state.
In 2010,  Wang \emph{et al}. \cite{Wang2010} proposed a practical scheme for concentrating $W$ state $\alpha|HHV\rangle+\beta(|HVH\rangle+|VHH\rangle)$ with linear optics. Here $|H\rangle$ and $|V\rangle$ denote the photons in  the horizontal and vertical linear polarizations states, respectively.
Yildiz \cite{Yildiz} proposed a scheme for distilling asymmetric $W$ states $\frac{1}{\sqrt{2}}|001\rangle+\frac{1}{2}|010\rangle+\frac{1}{2}|100\rangle$ and $\frac{1}{2}|001\rangle+\frac{1}{2}|010\rangle+\frac{1}{\sqrt{2}}|100\rangle$.
In 2012, Sheng \emph{et al}. \cite{Sheng2012w} presented linear-optics-based and cross-Kerr-based ECPs for $W$ states with known parameters.
Yan \emph{et al}. \cite{Yan2014} designed an ECP for four-photon cluster states, utilizing cross-Kerr nonlinearity and CNOT gate.
In 2015, Sheng \emph{et al}. \cite{Sheng2015w} used different parity check gates to concentrate $N$-particle $W$ states.
In 2017, Zhang \emph{et al}. \cite{Zhang2017}  proposed an ECP resorting to circuit quantum electrodynamics.
In recent years, much attention has been paid to hyperentanglement concentration protocols (hyper-ECPs) due to their excellent properties such as high capacity, low loss rate, and fewer experimental requirements \cite{hyper00,hyper01,hyper1,hyper2,hyper3,hyper4,split}.

Parameter-splitting is the current optimal strategy to implement ECP for less-entangled states with known parameters \cite{split}.
Post-selection principles are necessary for existing linear optical ECP for unknown less-entangled states \cite{Zhao,experiment1,experiment2,Sheng2012} as polarizing beam splitters (PBSs) are employed to pick up the desired instances in which each of the spatial contains exactly one photon. The destructive photon-number-resolving detectors can be used to  discard the case that the photon pair coincidence at one spatial.
However, such sophisticated detectors are not likely to be available with current technology, which makes the linear optical ECPs cannot be accomplished simply.
In addition, the recycling strategies are only introduced to increase the success probability of the cross-Kerr-based ECPs \cite{Sheng2008,Sheng2012}.
Hence, it is significant to investigate heralded and recyclable ECPs for partially entangled states without post-selection or photon-number-resolving detectors.

In this paper, we first present a heralded ECP for unknown less-entangled Bell states resorting to linear optical elements. Compared to the previous schemes, we avoid the need for photon-number-resolving detectors or post-selection principles by introducing the time-delay degree of freedom (DOF).
Our scheme is heralded unambiguously by the detection signatures, which makes our ECP much more practical.
The incident photons where distillation fails are kept in the less-entangled Bell state, and employing the recycling strategies can improve the success probability of concentration from 0.5 to 0.75 in  principle.
Only the probability of approaching the target state in an open quantum system can reach unity by iteration due to quantum anti-Zeno effect \cite{QAZE1,QAZE2}.
Moreover, the program is also available for ECP for multi-photon less-entangled unknown GHZ states, and the schemes are later designed in detail.
The presented architectures for ECPs with linear optics can be exactly realized with current experimental technology.


\section{Heralded ECP for unknown Bell state with linear optics}\label{sec2}

In the section, we present a heralded ECP for two-photon polarization less-entangled Bell states with unknown parameters using linear optics. By introducing the time-delay DOF to the detected photons, our ECP can be exactly heralded by the detection signatures.
The entanglement concentration process does not rely on post-selection principle.

\begin{figure} [htbp]
\begin{center}
\includegraphics[width=15cm]{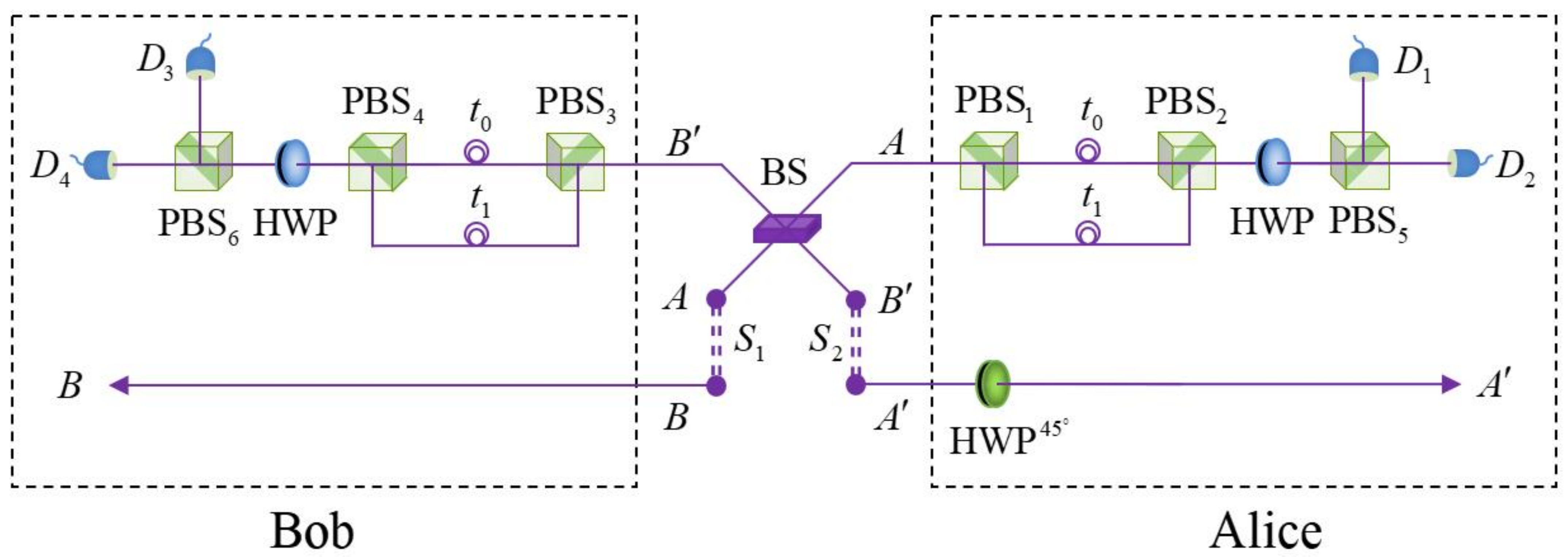}
\caption{Schematic diagram of the ECP for a partially entangled Bell state with unknown parameters.
  $S_{1}$ and $S_{2}$ are two pairs of identical entanglement sources for $|\phi\rangle_{AB}$ and $|\phi\rangle_{A^{\prime} B^{\prime}}$, respectively.
  $\textrm{BS}$ denotes the 50:50 beam splitter.
  $\textrm{PBS}_{i}$ $(i=1,2,\cdots,6)$ is a polarizing beam splitter which transmits the $H$-polarization component and reflects the $V$-polarization component.
  $\textrm{HWP}^{45^{\circ}}$ and $\textrm{HWP}$ represent half-wave plates oriented at $45^{\circ}$ and $22.5^{\circ}$, respectively.
  $D_{i}$ $(i=1,2,3,4)$ is a single-photon detector.
  The optical circle on the spatial mode denotes time delay $t_{0}$ or $t_{1}$.}\label{Fig.1}
\end{center}
\end{figure}

Suppose two maximally entangled Bell states $|\phi\rangle_{AB}$ and $|\phi\rangle_{A^{\prime}B^{\prime}}$ are generated initially from $S_{1}$ and $S_{2}$, respectively.
Here
\begin{eqnarray}\label{eq1}
|\phi\rangle_{AB}=\frac{1}{\sqrt{2}}(|HH\rangle+|VV\rangle)_{AB},\quad
|\phi\rangle_{A^{\prime}B^{\prime}}=\frac{1}{\sqrt{2}}(|HH\rangle+|VV\rangle)_{A^{\prime}B^{\prime}}.
\end{eqnarray}
The state of four-photon system composed of photons $A$, $B$, $A^{\prime}$, and $B^{\prime}$ can be described as
\begin{eqnarray}\label{eq2}
\begin{aligned}
|\Phi_{0}\rangle&=|\phi\rangle_{A B} \otimes |\phi\rangle_{A^{\prime} B^{\prime}}\\
    &=\frac{1}{2}(|HH\rangle+|VV\rangle)_{AB} \otimes (|HH\rangle+|VV\rangle)_{A^{\prime}B^{\prime}}.
\end{aligned}
\end{eqnarray}

Then as shown in Fig. \ref{Fig.1}, photons $A$ and $B^{\prime}$ immediately pass through a 50:50 beam splitter (BS), resulting in the following transformations
\begin{eqnarray}\label{eq3}
&|\Gamma\rangle_{A} \xrightarrow{\text{BS}} \frac{1}{\sqrt{2}}(|\Gamma\rangle_{A}+|\Gamma\rangle_{B^{\prime}}),
\quad
|\Gamma\rangle_{B^{\prime}} \xrightarrow{\text{BS}} \frac{1}{\sqrt{2}}(-|\Gamma\rangle_{A}+|\Gamma\rangle_{B^{\prime}}),
\end{eqnarray}
where $|\Gamma\rangle$ represents the polarization state $|H\rangle$ or $|V\rangle$.
Considering Eq. (\ref{eq3}), after passing the BS, $|\Phi_{0}\rangle$ is transformed into
\begin{eqnarray}\label{eq4}
\begin{aligned}
|\Phi_{1}\rangle=&\frac{1}{4}[(|H\rangle_{A} + |H\rangle_{B^{\prime}})|H\rangle_{B}
+ (|V\rangle_{A} + |V\rangle_{B^{\prime}})|V\rangle_{B}]\\
& \otimes  [|H\rangle_{A^{\prime}}(|H\rangle_{B^{\prime}}-|H\rangle_{A})
+ |V\rangle_{A^{\prime}}(|V\rangle_{B^{\prime}}-|V\rangle_{A})].
\end{aligned}
\end{eqnarray}
Subsequently, photons $A$ and $A^{\prime}$ ($B$ and $B^{\prime}$) of the state $|\Phi_{1}\rangle$ is sent to Alice (Bob), and owing to the
noisy channels, $|\Phi_{1}\rangle$ may decay to a partially less-entangled state
\begin{eqnarray}\label{eq5}
\begin{aligned}
|\Phi_{2}\rangle=&\frac{1}{2}[\alpha(|H\rangle_{A} + |H\rangle_{B^{\prime}})|H\rangle_{B} + \beta(|V\rangle_{A} + |V\rangle_{B^{\prime}})|V\rangle_{B}]\\
& \otimes  [\alpha|H\rangle_{A^{\prime}}(|H\rangle_{B^{\prime}}-|H\rangle_{A}) + \beta|V\rangle_{A^{\prime}}(|V\rangle_{B^{\prime}}-|V\rangle_{A})],
\end{aligned}
\end{eqnarray}
where the unknown parameters $\alpha$ and $\beta$ satisfy the normalization relation $|\alpha|^{2}+|\beta|^{2}=1$.

In order to distill the maximally entangled Bell state $(|HH\rangle+|VV\rangle)/\sqrt{2}$ from $|\Phi_{2}\rangle$, the two distant parties, Alice and Bob, need to complete the operations shown in Fig. \ref{Fig.1}.
To describe this process more clearly, combined with the Hong-Ou-Mandel effect, we rewrite Eq. (\ref{eq5}) in the following normalized form
\begin{eqnarray}\label{eq6}
\begin{aligned}
|\Phi_{2}\rangle=&\frac{1}{\sqrt{2}}[\alpha^2|HH\rangle_{A^{\prime}B}(|HH\rangle_{B^{\prime}B^{\prime}} - |HH\rangle_{AA}) + \beta^2|VV\rangle_{A^{\prime}B}(|VV\rangle_{B^{\prime}B^{\prime}} - |VV\rangle_{AA})]\\
& + \frac{1}{2}[\alpha\beta(|VH\rangle+|HV\rangle)_{A^{\prime}B}|HV\rangle_{B^{\prime}B^{\prime}} - \alpha\beta(|VH\rangle+|HV\rangle)_{A^{\prime}B}|HV\rangle_{AA}\\
& + \alpha\beta(|VH\rangle-|HV\rangle)_{A^{\prime}B}|HV\rangle_{AB^{\prime}} - \alpha\beta(|VH\rangle-|HV\rangle)_{A^{\prime}B}|VH\rangle_{AB^{\prime}}].
\end{aligned}
\end{eqnarray}
Specifically, Alice flips the state of photon $A^{\prime}$ by using a half-wave plate oriented at $45^{\circ}$. That is, $\textrm{HWP}^{45^{\circ}}$ completes the transformations $|H\rangle \xrightarrow{\text{HWP}^{45^\circ}} |V\rangle$ and $|V\rangle \xrightarrow{\text{HWP}^{45^\circ}} |H\rangle$.
Hence, $\textrm{HWP}^{45^{\circ}}$ transforms $|\Phi_{2}\rangle$ into
\begin{eqnarray}\label{eq7}
\begin{aligned}
|\Phi_{3}\rangle=&\frac{1}{\sqrt{2}}[\alpha^2|VH\rangle_{A^{\prime}B}(|HH\rangle_{B^{\prime}B^{\prime}} - |HH\rangle_{AA}) + \beta^2|HV\rangle_{A^{\prime}B}(|VV\rangle_{B^{\prime}B^{\prime}} - |VV\rangle_{AA})]\\
& + \frac{1}{2}[\alpha\beta(|HH\rangle+|VV\rangle)_{A^{\prime}B}|HV\rangle_{B^{\prime}B^{\prime}} - \alpha\beta(|HH\rangle+|VV\rangle)_{A^{\prime}B}|HV\rangle_{AA}\\
& + \alpha\beta(|HH\rangle-|VV\rangle)_{A^{\prime}B}|HV\rangle_{AB^{\prime}} - \alpha\beta(|HH\rangle-|VV\rangle)_{A^{\prime}B}|VH\rangle_{AB^{\prime}}].
\end{aligned}
\end{eqnarray}

Nextly, by using an unbalanced interferometer consisting of two PBSs,  Alice (Bob) introduces time-delays $t_0$ and $t_1$ to the $H$- and  $V$-polarization components of photon $A$ ($B^{\prime}$), respectively. Here $t_{0}$ and $t_{1}$ satisfy $\omega(t_{0}-t_{1})=2n\pi$, where $n$ is the nonzero integer. And then, the state $|\Phi_{3}\rangle$ becomes
\begin{eqnarray}\label{eq8}
\begin{aligned}
|\Phi_{4}\rangle=&\frac{1}{\sqrt{2}}[\alpha^2|VH\rangle_{A^{\prime}B}(|H_{t_{0}}H_{t_{0}}\rangle_{B^{\prime}B^{\prime}} - |H_{t_{0}}H_{t_{0}}\rangle_{AA})\\
& + \beta^2|HV\rangle_{A^{\prime}B}(|V_{t_{1}}V_{t_{1}}\rangle_{B^{\prime}B^{\prime}} - |V_{t_{1}}V_{t_{1}}\rangle_{AA})]\\
& + \frac{1}{2}[\alpha\beta(|HH\rangle+|VV\rangle)_{A^{\prime}B}|H_{t_{0}}V_{t_{1}}\rangle_{B^{\prime}B^{\prime}}\\
&-\alpha\beta(|HH\rangle+|VV\rangle)_{A^{\prime}B}|H_{t_{0}}V_{t_{1}}\rangle_{AA}\\
&+\alpha\beta(|HH\rangle-|VV\rangle)_{A^{\prime}B}|H_{t_{0}}V_{t_{1}}\rangle_{AB^{\prime}}\\
&-\alpha\beta(|HH\rangle-|VV\rangle)_{A^{\prime}B}|V_{t_{1}}H_{t_{0}}\rangle_{AB^{\prime}}].
\end{aligned}
\end{eqnarray}

After that, Alice (Bob) performs a Hadamard operation on photon $A$ ($B^{\prime}$) with a half wave plate ($\textrm{HWP}$). That is, $\textrm{HWP}$ completes the transformations
\begin{eqnarray}\label{eq9}
&|H\rangle \xrightarrow{\text{HWP}} \frac{1}{\sqrt{2}}(|H\rangle+|V\rangle),\quad
|V\rangle \xrightarrow{\text{HWP}} \frac{1}{\sqrt{2}}(|H\rangle-|V\rangle).
\end{eqnarray}
Those rotations convert $|\Phi_{4}\rangle$ into
\begin{eqnarray}\label{eq10}
\begin{aligned}
|\Phi_{5}\rangle
=&\frac{\sqrt{|\alpha|^{4}+|\beta|^{4}}}{2\sqrt{2}}|\phi_{1}^+\rangle_{A^{\prime}B}
[D_{B^{\prime} B^{\prime}}(0)(|HH\rangle
+|VV\rangle)_{B^{\prime} B^{\prime}}
-D_{A A}(0)(|HH\rangle+|VV\rangle)_{A A}]\\&
+\frac{\sqrt{|\alpha|^{4}+|\beta|^{4}}}{2\sqrt{2}}|\phi_{1}^-\rangle_{A^{\prime}B}
[D_{B^{\prime} B^{\prime}}(0)(|HV\rangle
+|VH\rangle)_{B^{\prime} B^{\prime}}
-D_{A A}(0)(|HV\rangle+|VH\rangle)_{A A}]\\&
+\frac{\alpha\beta}{2\sqrt{2}}|\phi^+\rangle_{A^{\prime}B}
[(|H_{t_{0}}H_{t_{1}}\rangle
-|H_{t_{0}}V_{t_{1}}\rangle
+|V_{t_{0}}H_{t_{1}}\rangle
-|V_{t_{0}}V_{t_{1}}\rangle)_{B^{\prime}B^{\prime}}
-(|H_{t_{0}}H_{t_{1}}\rangle\\&
-|H_{t_{0}}V_{t_{1}}\rangle
+|V_{t_{0}}H_{t_{1}}\rangle
-|V_{t_{0}}V_{t_{1}}\rangle)_{AA}]
+\frac{\alpha\beta}{2\sqrt{2}}|\phi^-\rangle_{A^{\prime}B}
[(|H_{t_{0}}H_{t_{1}}\rangle
-|H_{t_{0}}V_{t_{1}}\rangle\\&
+|V_{t_{0}}H_{t_{1}}\rangle
-|V_{t_{0}}V_{t_{1}}\rangle
-|H_{t_{1}}H_{t_{0}}\rangle
-|H_{t_{1}}V_{t_{0}}\rangle
+|V_{t_{1}}H_{t_{0}}\rangle
+|V_{t_{1}}V_{t_{0}}\rangle)_{AB^{\prime}}].
\end{aligned}
\end{eqnarray}
where
\begin{eqnarray}\label{eq11}
|\phi^\pm\rangle_{A^{\prime}B}=\frac{1}{\sqrt{2}}(|HH\rangle \pm |VV\rangle)_{A^{\prime}B},
\end{eqnarray}
\begin{eqnarray}\label{eq12}
|\phi_{1}^\pm\rangle_{A^{\prime}B}=(\alpha^{\prime}|VH\rangle \pm \beta^{\prime}|HV\rangle)_{A^{\prime}B}
\end{eqnarray}
with $\alpha^{\prime}=\frac{\alpha^{2}}{\sqrt{|\alpha|^{4}+|\beta|^{4}}}$, $\beta^{\prime}=\frac{\beta^{2}}{\sqrt{|\alpha|^{4}+|\beta|^{4}}}$.
Here $D_{AA(B^{\prime}B^{\prime})}(0)$ represents that there is no relative time-delay between two photons $A$ ($B^{\prime}$).
That is, there is no time interval between the reaction of the single-photon detector held by Alice or the single-photon detector held by Bob.

Finally, Alice (Bob) uses PBS$_{5}$ (PBS$_{6}$) and two single-photon detectors $\{D_{1}, D_{2}\}$  ($\{D_{3}, D_{4}\}$) to complete the measurement on the outing
photon $A$ $(B^{\prime})$ in the basis $\{|H\rangle, |V\rangle\}$.
The relationship between the detection signatures, the corresponding output states, and the feed-forward operations on photon $B$ is given in Tab. \ref{Table1}.
If detector pair $(D_i, D_j)$ $(i,j=1,2,3,4)$ triggers with a time interval of $|t_{0}-t_{1}|$, they will get the desired maximally entangled state
$|\phi^+\rangle_{A^{\prime}B}$ with a success probability of $P=2|\alpha\beta|^2$ after applying the corresponding feed-forward operation shown in Tab. \ref{Table1}.
Otherwise, it means that detector pair $(D_i, D_j)$ fires without time interval. In such case, performing the feed-forward operation, they can get the normalization state $|\phi_{1}^{+}\rangle_{A^{\prime}B}$ with a probability of $|\alpha|^4+|\beta|^4 =1-2|\alpha\beta|^2$.

\begin{table}[htbp]
  \centering
  \caption{The relations between the detection signatures and the classical feed-forward operations to complete the ECP for Bell state. The operation $\sigma_{z}=|H\rangle \langle H|-|V\rangle \langle V| $ can be accomplished with a half-wave plate oriented at $0^\circ$. $(D_{i}^{t_{0}},D_{j}^{t_{1}})$ or $(D_{i}^{t_{1}},D_{j}^{t_{0}})$ $(i,j=1,2,3,4)$ indicates that $(D_i, D_j)$ triggers with a time interval of $|t_{0}-t_{1}|$.}\label{Table1}
  \begin{tabular}{cccc}
  \hline   \hline
Single-photon  & Outcomes  & Feed-forward   & Success  \\
detectors  & of $A^{\prime}$ and $B$  & on $B$   &    probability \\ \hline
 $D_{1}, D_{2}, D_{3}, D_{4}$ & $|\phi_{1}^{+}\rangle_{A^{\prime}B}$  & none & \multirow{2}{*}{$|\alpha|^4+|\beta|^4$} \\ 
 $(D_{1},D_{2}), (D_{3},D_{4})$ & $|\phi_{1}^{-}\rangle_{A^{\prime}B}$ & $\sigma_{z}$ &\\
\hline
 $(D_{1}^{t_{0}},D_{1}^{t_{1}}), (D_{2}^{t_{0}},D_{2}^{t_{1}}),(D_{3}^{t_{0}},D_{3}^{t_{1}}),(D_{4}^{t_{0}},D_{4}^{t_{1}})$ & \multirow{2}{*}{$|\phi^{+}\rangle_{A^{\prime}B}$}  & \multirow{2}{*}{none} & \multirow{4}{*}{$2|\alpha\beta|^2$}\\ 
$(D_{1}^{t_{0}},D_{2}^{t_{1}}), (D_{1}^{t_{1}},D_{2}^{t_{0}}),(D_{3}^{t_{0}},D_{4}^{t_{1}}), (D_{3}^{t_{1}},D_{4}^{t_{0}})$ &  &   &\\ 
$(D_{1}^{t_{0}},D_{3}^{t_{1}}), (D_{1}^{t_{1}},D_{3}^{t_{0}}),(D_{2}^{t_{0}},D_{4}^{t_{1}}), (D_{2}^{t_{1}},D_{4}^{t_{0}})$ & \multirow{2}{*}{$|\phi^{-}\rangle_{A^{\prime}B}$} & \multirow{2}{*}{$\sigma_{z}$}   &\\ 
$(D_{2}^{t_{0}},D_{3}^{t_{1}}), (D_{2}^{t_{1}},D_{3}^{t_{0}}),(D_{1}^{t_{0}},D_{4}^{t_{1}}), (D_{1}^{t_{1}},D_{4}^{t_{0}})$ &  &  &\\
\hline  \hline
\end{tabular}
\end{table}

\begin{figure} [htbp]
  \centering
  \includegraphics[width=9cm]{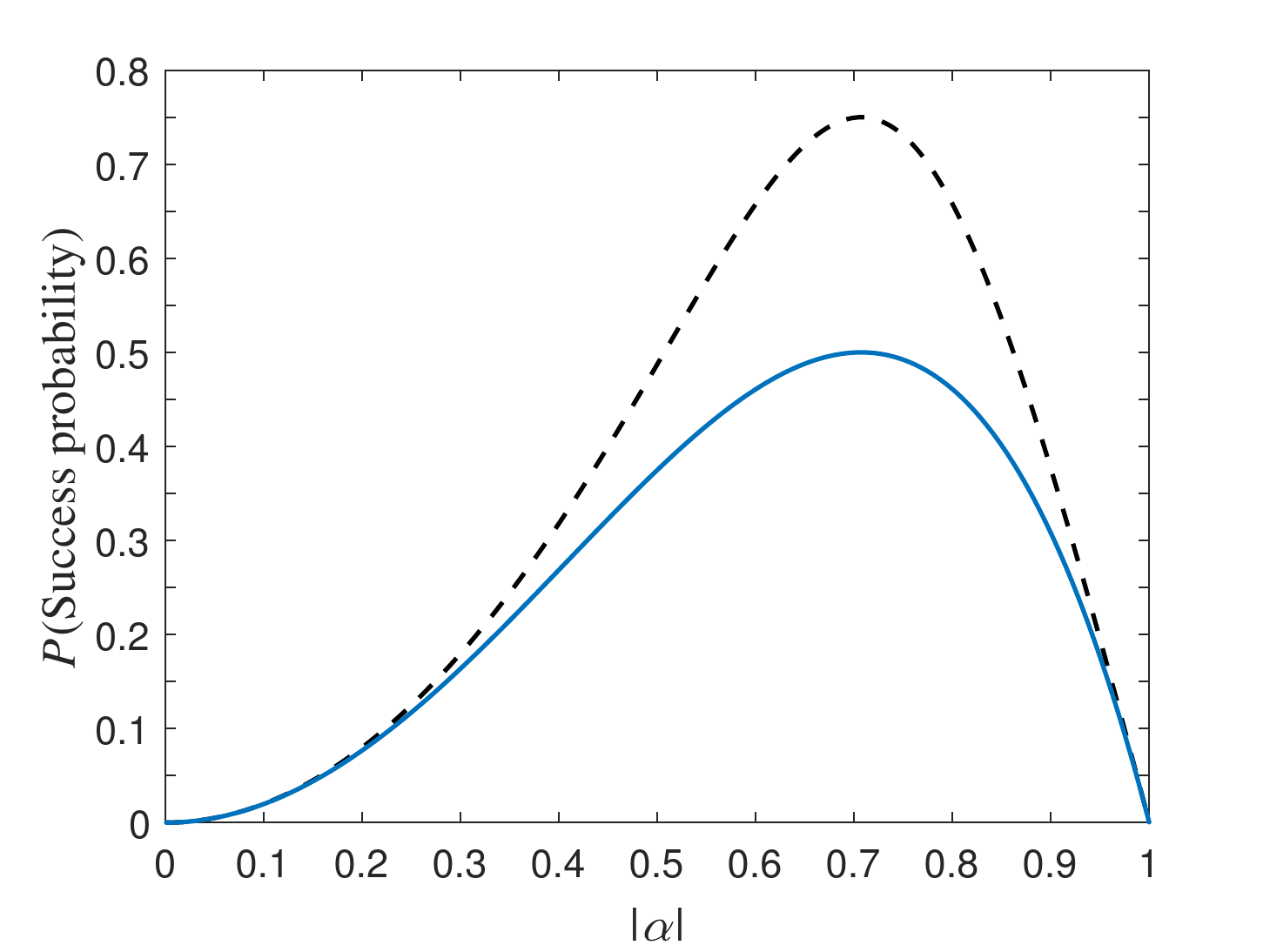}
  \caption{The success probability of the ECP for Bell states as a function of parameter $\alpha$. Here $|\alpha|=\sqrt{1-|\beta|^2}\in(0,1)$. The dotted line depicts the total success rate after recycling $|\phi_{1}^+\rangle_{A^{\prime}B}$.}\label{Fig.2}
\end{figure}

The solid line in Fig. \ref{Fig.2} shows the success probability of the presented ECP, and $|\alpha|=\sqrt{1-|\beta|^2}\in(0,1)$ are taken.
Besides, it is obvious that the photons $A^{\prime}$ and $B$ kept are in the state $|\phi^+\rangle_{A^{\prime}B}$ or $|\phi_{1}^+\rangle_{A^{\prime}B}$. Alice and Bob can further distill $|\phi^+\rangle_{A^{\prime}B}$ from $|\phi_{1}^+\rangle_{A^{\prime}B}$, because $|\phi_{1}^+\rangle_{A^{\prime}B}$ passing through a $\textrm{HWP}^{45^{\circ}}$ for photons $A^{\prime}$ has the similar form as $|\phi\rangle_{AB}$ subjected to channel noise \cite{Zhao,Sheng2008}.
Therefore, by recycling the state $|\phi_{1}^+\rangle_{A^{\prime}B}$ and applying the ECP of Ref. \cite{Zhao} or \cite{Sheng2012}, the success probability can be efficiently improved by
$(|\alpha|^{4}+|\beta|^{4})\cdot2|\alpha^{\prime}\beta^{\prime}|^{2}
=\frac{2|\alpha\beta|^{4}}{|\alpha|^{4}+|\beta|^{4}}$.
As the dotted line in Fig. \ref{Fig.2} shows, the total success probability has increased from 0.5 to 0.75 in principle.

\section{Heralded ECP for unknown GHZ state with linear optics}\label{sec3}

Our heralded ECP for polarization unknown Bell states with linear optics can be generalized to the case of multi-photon GHZ states.
Suppose two maximally entangled GHZ states $|\psi\rangle_{ABC}$ and $|\psi\rangle_{A^{\prime}B^{\prime}C^{\prime}}$ are generated initially from $S_{1}$ and $S_{2}$, respectively.
Here
\begin{eqnarray}\label{eq13}
|\psi\rangle_{ABC}=\frac{1}{\sqrt{2}}(|HHH\rangle+|VVV\rangle)_{ABC},\,\,\,\,
|\psi\rangle_{A^{\prime}B^{\prime}C^{\prime}}=\frac{1}{\sqrt{2}}(|HHH\rangle+|VVV\rangle)_{A^{\prime}B^{\prime}C^{\prime}}.
\end{eqnarray}
The state of the six-photon system composed of photons $A$, $B$, $C$, $A^{\prime}$, $B^{\prime}$, and $C^{\prime}$ is given by
\begin{eqnarray}\label{eq14}
\begin{aligned}
|\Psi_{0}\rangle=&|\psi\rangle_{A B} \otimes |\psi\rangle_{A^{\prime} B^{\prime}}\\
    =&\frac{1}{2}(|HHH\rangle+|VVV\rangle)_{ABC} \otimes (|HHH\rangle+|VVV\rangle)_{A^{\prime}B^{\prime}C^{\prime}}.
\end{aligned}
\end{eqnarray}

\begin{figure}  [htbp]
  \centering
  \includegraphics[width=9cm]{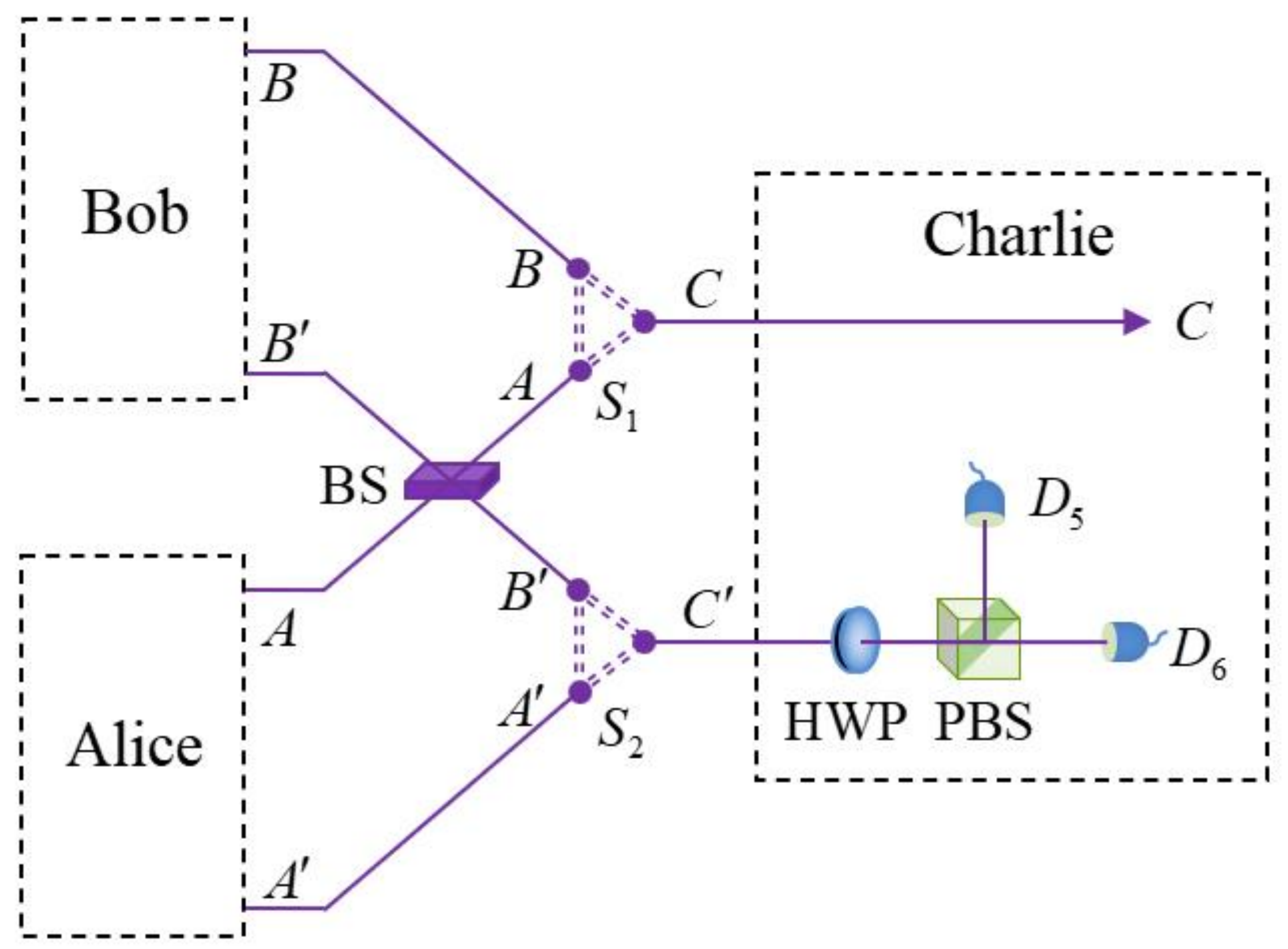}
  \caption{Schematic diagram of the ECP for a three-photon GHZ states with unknown parameters.  $S_{1}$ and $S_{2}$ are entanglement sources for $|\psi\rangle_{ABC}$ and $|\psi\rangle_{A^{\prime} B^{\prime} C^{\prime}}$, respectively. The setups in dashed boxes held by Alice and Bob are shown in Fig. \ref{Fig.1}.}\label{Fig.3}
\end{figure}

As shown in Fig. \ref{Fig.3}, photons $A$ and $B^{\prime}$ pass through a BS, and then photons pairs $AA^{\prime}$, $BB^{\prime}$, and $CC^{\prime}$ of the state $|\Psi_{0}\rangle$ is sent to three distant parties Alice, Bob, and Charlie, respectively.
Owing to the noisy channels, $|\Psi_{0}\rangle$ may decay to a partially less-entangled state
\begin{eqnarray}\label{eq15}
\begin{aligned}
|\Psi_{1}\rangle=&\frac{1}{2}[\alpha(|H\rangle_{A} + |H\rangle_{B^{\prime}})|H\rangle_{B}|H\rangle_{C} + \beta(|V\rangle_{A} + |V\rangle_{B^{\prime}})|V\rangle_{B}|V\rangle_{C}]\\
& \otimes  [\alpha|H\rangle_{A^{\prime}}(|H\rangle_{B^{\prime}}-|H\rangle_{A})|H\rangle_{C^{\prime}} + \beta|V\rangle_{A^{\prime}}(|V\rangle_{B^{\prime}}-|V\rangle_{A})|V\rangle_{C^{\prime}}]
\end{aligned}
\end{eqnarray}
where the unknown parameters $\alpha$ and $\beta$ satisfy the normalization relation $|\alpha|^{2}+|\beta|^{2}=1$.

The dashed boxes in Fig. \ref{Fig.3} held by Alice and Bob are the same setups shown in Fig. \ref{Fig.1}.
To be specific,  Alice performs the $\sigma_x$ operation on photon $A^{\prime}$. After executing this operation, time delays $t_0$ and $t_1$ are introduced by Alice (Bob) to photons $A$ ($B^{\prime}$) by using the balanced interferometers, i.e.,
$|H\rangle_{A} \rightarrow |H_{t_0}\rangle_{A}$,
$|V\rangle_{A} \rightarrow |V_{t_1}\rangle_{A}$
($|H\rangle_{B^{\prime}} \rightarrow |H_{t_0}\rangle_{B^{\prime}}$,
 $|V\rangle_{B^{\prime}} \rightarrow |V_{t_1}\rangle_{B^{\prime}}$).
Then, $|\Psi_{1}\rangle$ is converted to
\begin{eqnarray}\label{eq16}
\begin{aligned}
|\Psi_{2}\rangle=&\frac{1}{\sqrt{2}}[\alpha^2|VHHH\rangle_{A^{\prime}BCC^{\prime}}(|H_{t_0}H_{t_0}\rangle_{B^{\prime}B^{\prime}} - |H_{t_0}H_{t_0}\rangle_{AA})\\
&+ \beta^2|HVVV\rangle_{A^{\prime}BCC^{\prime}}(|V_{t_1}V_{t_1}\rangle_{B^{\prime}B^{\prime}} - |V_{t_1}V_{t_1}\rangle_{AA})]\\&
+ \frac{\alpha\beta}{2}[(|HHHV\rangle+|VVVH\rangle)_{A^{\prime}BCC^{\prime}}|H_{t_0}V_{t_1}\rangle_{B^{\prime}B^{\prime}}\\&
- (|HHHV\rangle+|VVVH\rangle)_{A^{\prime}BCC^{\prime}}|H_{t_0}V_{t_1}\rangle_{AA}\\&
+ (|HHHV\rangle-|VVVH\rangle)_{A^{\prime}BCC^{\prime}}|H_{t_0}V_{t_1}\rangle_{AB^{\prime}}\\&
- (|HHHV\rangle-|VVVH\rangle)_{A^{\prime}BCC^{\prime}}|V_{t_1}H_{t_0}\rangle_{AB^{\prime}}].
\end{aligned}
\end{eqnarray}

Then, as shown in Fig. \ref{Fig.3}, Alice, Bob, and Charlie lead photons $A$, $B^{\prime}$, and $C^{\prime}$ to pass through $\textrm{HWP}$, respectively. These half-plate waves transform $|\Psi_{2}\rangle$ into
\begin{eqnarray}\label{eq17}
\begin{aligned}
|\Psi_{3}\rangle
=&\frac{\sqrt{|\alpha|^{4}+|\beta|^{4}}}{2\sqrt{2}}|\psi_{1}^+\rangle_{A^{\prime}BC}
[D_{B^{\prime} B^{\prime}}(0)(|HHH\rangle+|HVV\rangle
+|VHV\rangle+|VVH\rangle)_{B^{\prime} B^{\prime} C^{\prime}}\\&
-D_{A A}(0)(|HHH\rangle+|HVV\rangle
+|VHV\rangle+|VVH\rangle)_{A A C^{\prime}}]\\&
+\frac{\sqrt{|\alpha|^{4}+|\beta|^{4}}}{2\sqrt{2}}|\psi_{1}^-\rangle_{A^{\prime}BC}
[D_{B^{\prime} B^{\prime}}(0)(|HHV\rangle+|HVH\rangle+|VHH\rangle+|VVV\rangle)_{B^{\prime} B^{\prime} C^{\prime}}\\&
-D_{A A}(0)(|HHV\rangle+|HVH\rangle+|VHH\rangle+|VVV\rangle)_{AA C^{\prime}}]\\&
+\frac{\alpha\beta}{2}|\psi^{+}\rangle_{A^{\prime}BC}
[(|H_{t_{0}} H_{t_{1}} H\rangle
-|H_{t_{0}} V_{t_{1}} H\rangle
+|V_{t_{0}} H_{t_{1}} H\rangle
-|V_{t_{0}} V_{t_{1}} H\rangle)_{B^{\prime}B^{\prime}C^{\prime}}\\&
-(|H_{t_{0}} H_{t_{1}} H\rangle
-|H_{t_{0}} V_{t_{1}} H\rangle
+|V_{t_{0}} H_{t_{1}} H\rangle
-|V_{t_{0}} V_{t_{1}} H\rangle)_{AAC^{\prime}}\\&
-(|H_{t_{0}} H_{t_{1}} V\rangle
-|H_{t_{0}} V_{t_{1}} V\rangle
+|V_{t_{0}} H_{t_{1}} V\rangle
-|V_{t_{0}} V_{t_{1}} V\rangle)_{AB^{\prime}C^{\prime}}\\&
+(|H_{t_{1}} H_{t_{0}} V\rangle
+|H_{t_{1}} V_{t_{0}} V\rangle
-|V_{t_{1}} H_{t_{0}} V\rangle
-|V_{t_{1}} V_{t_{0}} V\rangle)_{AB^{\prime}C^{\prime}}]\\&
+\frac{\alpha\beta}{2}|\psi^{-}\rangle_{A^{\prime}BC}
[(|H_{t_{0}} H_{t_{1}} V\rangle
-|H_{t_{0}} V_{t_{1}} V\rangle
+|V_{t_{0}} H_{t_{1}} V\rangle
-|V_{t_{0}} V_{t_{1}} V\rangle)_{AAC^{\prime}}\\&
-(|H_{t_{0}} H_{t_{1}} V\rangle
-|H_{t_{0}} V_{t_{1}} V\rangle
+|V_{t_{0}} H_{t_{1}} V\rangle
-|V_{t_{0}} V_{t_{1}} V\rangle)_{B^{\prime}B^{\prime}C^{\prime}}\\&
+(|H_{t_{0}} H_{t_{1}} H\rangle
-|H_{t_{0}} V_{t_{1}} H\rangle
+|V_{t_{0}} H_{t_{1}} H\rangle
-|V_{t_{0}} V_{t_{1}} H\rangle)_{AB^{\prime}C^{\prime}}\\&
-(|H_{t_{1}} H_{t_{0}} H\rangle
+|H_{t_{1}} V_{t_{0}} H\rangle
-|V_{t_{1}} H_{t_{0}} H\rangle
-|V_{t_{1}} V_{t_{0}}  H\rangle)_{AB^{\prime}C^{\prime}}].
\end{aligned}
\end{eqnarray}

where
\begin{eqnarray}\label{eq18}
|\psi^\pm\rangle_{A^{\prime}BC}=\frac{1}{\sqrt{2}}(|HHH\rangle \pm |VVV\rangle)_{A^{\prime}BC},
\end{eqnarray}
\begin{eqnarray}\label{eq19}
|\psi_{1}^{\pm}\rangle_{A^{\prime}BC}=\frac{\alpha^{2}}{\sqrt{|\alpha|^{4}+|\beta|^{4}}}|VHH\rangle_{A^{\prime}BC} \pm \frac{\beta^{2}}{\sqrt{|\alpha|^{4}+|\beta|^{4}}}|HVV\rangle_{A^{\prime}BC}.
\end{eqnarray}

Finally, the outcomes of photons $A$, $B^{\prime}$, and $C^{\prime}$ are measured by PBSs and single-photon detectors.
Tab. \ref{Table2} depicts the detection signatures and the output states.
When the detectors held by Alice and Bob are triggered with a time interval of $|t_{0}-t_{1}|$,  Alice, Bob, and  Charlie will get
the desired maximally entangled state $|\psi^+\rangle_{A^{\prime}BC}$ with a success probability of $2|\alpha\beta|^{2}$, after performing the feed-forward operations on photon $B$.
If the detectors held by Alice and Bob are triggered simultaneously, they will get the recyclable normalization state $|\psi_{1}^{+}\rangle_{A^{\prime}BC}$ with a probability of $|\alpha|^{4}+|\beta|^{4}$.
The same argument as that made in Section \ref{sec2}, the success probability can be increased by recycling $|\psi_{1}^{+}\rangle_{A^{\prime}BC}$ and that equals to the dotted line in Fig. \ref{Fig.2}.

\begin{table}[htbp]
  \centering
  \caption{The relations between the detection signatures and the classical feed-forward operations to complete the ECP for GHZ states.
  }\label{Table2}
  \begin{tabular}{cccc}
  \hline   \hline
     Single- photon   & Outcomes     & Feed-forward & Success   \\
      detectors    & of $A^{\prime}$, $B$, and $C$   &    on $B$ & probability  \\ \hline
 $(D_{1},D_{2},D_{5}), (D_{3},D_{4},D_{5})$ & \multirow{2}{*}{$|\psi_{1}^{+}\rangle_{A^{\prime}BC}$}  & \multirow{2}{*}{none} & \multirow{4}{*}{$|\alpha|^4+|\beta|^4$} \\ 
     $(D_{1},D_{6}), (D_{2},D_{6}),(D_{3},D_{6}), (D_{4},D_{6})$ &  &  &  \\
     $(D_{1},D_{2},D_{6}), (D_{3},D_{4},D_{6})$
      & \multirow{2}{*}{$|\psi_{1}^{-}\rangle_{A^{\prime}BC}$} & \multirow{2}{*}{$\sigma_{z}$} &\\
     $(D_{1},D_{5}), (D_{2},D_{5}),(D_{3},D_{5}), (D_{4},D_{5})$ &  & &   \\ 
\hline
 $(D_{1}^{t_{0}},D_{1}^{t_{1}},D_{6}), (D_{2}^{t_{0}},D_{2}^{t_{1}},D_{6})$ & \multirow{8}{*}{$|\psi^{+}\rangle_{A^{\prime}BC}$} & \multirow{8}{*}{none} & \multirow{16}{*}{$2|\alpha\beta|^2$} \\
       $(D_{3}^{t_{0}},D_{3}^{t_{1}},D_{6}), (D_{4}^{t_{0}},D_{4}^{t_{1}},D_{6})$ &  & & \\
       $(D_{1}^{t_{0}},D_{2}^{t_{1}},D_{6}), (D_{1}^{t_{1}},D_{2}^{t_{0}},D_{6})$ &  & & \\ 
       $(D_{3}^{t_{0}},D_{4}^{t_{1}},D_{6}), (D_{3}^{t_{1}},D_{4}^{t_{0}},D_{6})$ &  & & \\
       $(D_{1}^{t_{0}},D_{3}^{t_{1}},D_{5}), (D_{1}^{t_{1}},D_{3}^{t_{0}},D_{5})$ &  & & \\ 
       $(D_{2}^{t_{0}},D_{4}^{t_{1}},D_{5}), (D_{2}^{t_{1}},D_{4}^{t_{0}},D_{5})$ &  & & \\ 
       $(D_{2}^{t_{0}},D_{3}^{t_{1}},D_{5}), (D_{2}^{t_{1}},D_{3}^{t_{0}},D_{5})$ &  & & \\
       $(D_{1}^{t_{0}},D_{4}^{t_{1}},D_{5}), (D_{1}^{t_{1}},D_{4}^{t_{0}},D_{5})$ &  & & \\\cline{1-3}
 $(D_{1}^{t_{0}},D_{1}^{t_{1}},D_{5}), (D_{2}^{t_{0}},D_{2}^{t_{1}},D_{5})$ & \multirow{8}{*}{$|\psi^{-}\rangle_{A^{\prime}BC}$} & \multirow{8}{*}{$\sigma_{z}$} \\
                    $(D_{3}^{t_{0}},D_{3}^{t_{1}},D_{5}), (D_{4}^{t_{0}},D_{4}^{t_{1}},D_{5})$ &  &  \\
                    $(D_{1}^{t_{0}},D_{2}^{t_{1}},D_{5}), (D_{1}^{t_{1}},D_{2}^{t_{0}},D_{5})$ &  &  \\ 
                    $(D_{3}^{t_{0}},D_{4}^{t_{1}},D_{5}), (D_{3}^{t_{1}},D_{4}^{t_{0}},D_{5})$ &  &  \\
                    $(D_{1}^{t_{0}},D_{3}^{t_{1}},D_{6}), (D_{1}^{t_{1}},D_{3}^{t_{0}},D_{6})$ &  &  \\ 
                    $(D_{2}^{t_{0}},D_{4}^{t_{1}},D_{6}), (D_{2}^{t_{1}},D_{4}^{t_{0}},D_{6})$ &  &  \\ 
                    $(D_{2}^{t_{0}},D_{3}^{t_{1}},D_{6}), (D_{2}^{t_{1}},D_{3}^{t_{0}},D_{6})$ &  &  \\
                    $(D_{1}^{t_{0}},D_{4}^{t_{1}},D_{6}), (D_{1}^{t_{1}},D_{4}^{t_{0}},D_{6})$ &  &  \\
\hline  \hline
\end{tabular}
\end{table}

\begin{figure} [htbp]
  \centering
  \includegraphics[width=8cm]{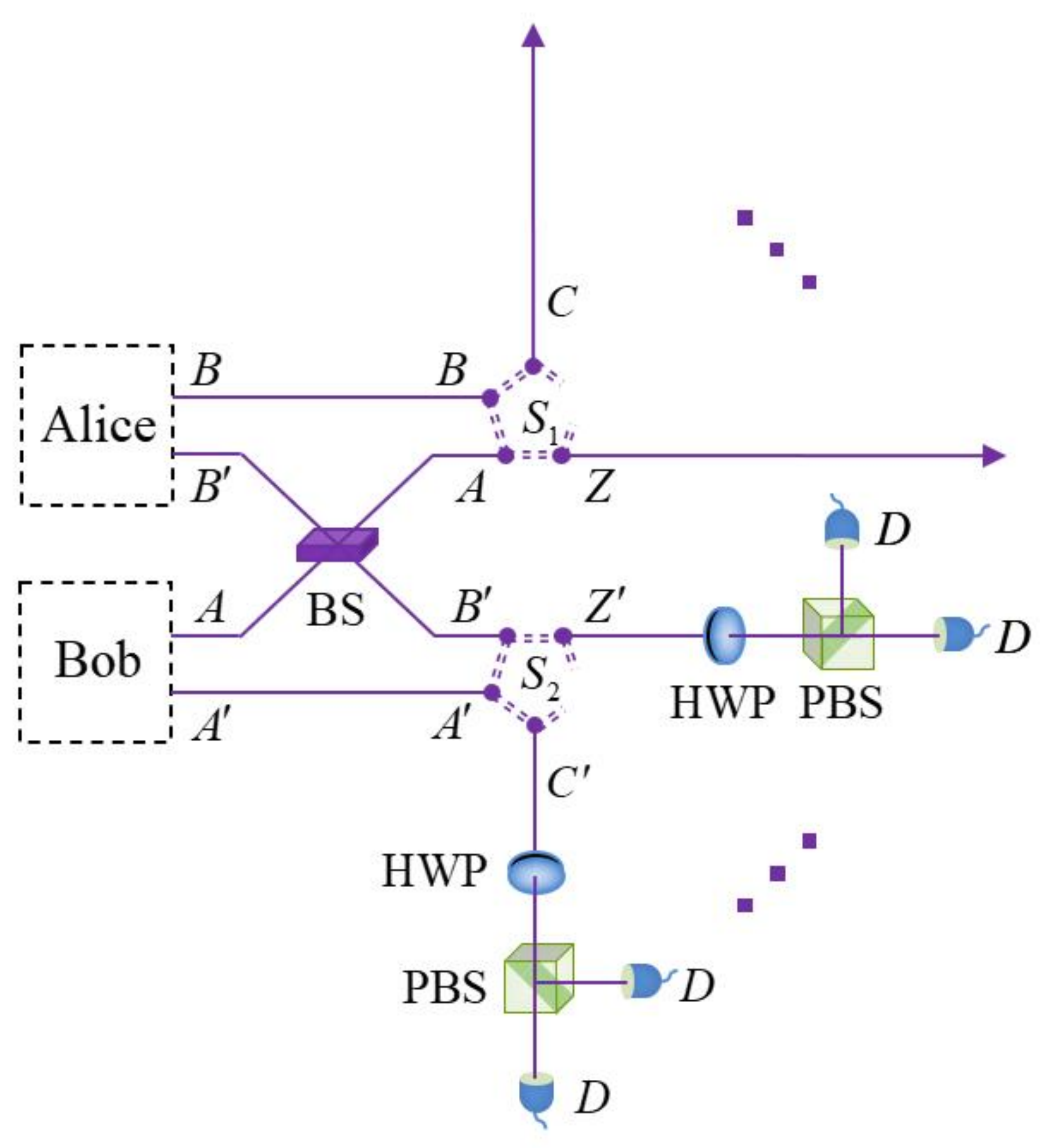}
  \caption{Schematic diagram of the  ECP for an arbitrary multi-photon GHZ state with unknown parameters. $S_{1}$ and $S_{2}$ are entanglement sources for two  multi-photon GHZ states. The operations on photons $C\cdots Z$ ($C^{\prime}\cdots Z^{\prime}$) are identical.}\label{Fig.4}
\end{figure}

The schematic setup to implement heralded and non-postselection ECP for  multi-photon GHZ states with unknown parameters is shown in Fig. \ref{Fig.4}. The ECP has the same success probability as the presented ECP for Bell states , and can also be increased to 0.75.

\section{Discussion and summary}\label{sec4}

Entanglement concentration is a powerful way to deal with the negative effects of environmental noise in the long-distance quantum communication.
The ECP with unknown parameters is more efficient than the one with known parameters as the environmental decoherence causes the parties blind to the information about the state.
However, the existing ECP for an unknown less-entangled state cannot be accomplished simply with linear optics as PBSs cannot exactly complete parity-check measurement without photon-number-resolving detectors.
Moreover, the destructive measurements, involving the photon pair coincidence at a detector, make the recycling strategy impossible.
Cross-Kerr medium and matter platforms such as atom, quantum dot, and nitrogen-vacancy color centre in diamonds have been employed to bypass the destructive measurements and photon-number-resolving detectors, and to increase the success probability. However, the giant Kerr nonlinearities are difficult to achieve in the experiment with current technology, and $\theta\approx 10^{-18}$ rad for natural Kerr media is extremely small. Implementations and manipulations of matter-qubit are challenged in experiment by inefficiency and impracticality.

The key element of the presented schemes is the time delay.
By introducing time-delay DOF to the photons $A$ and $B^{\prime}$, the detection signatures can exactly distinguish the even-parity state $\{|H_{t_{0}}H_{t_{0}}\rangle_{AA}, |V_{t_{1}}V_{t_{1}}\rangle_{AA}, |H_{t_{0}}H_{t_{0}}\rangle_{B^{\prime}B^{\prime}}, |V_{t_{1}}V_{t_{1}}\rangle_{B^{\prime}B^{\prime}}\}$ from the odd-parity state $\{|H_{t_{0}}V_{t_{1}}\rangle_{AB^{\prime}},$ $|V_{t_{1}}H_{t_{0}}\rangle_{AB^{\prime}}, |H_{t_{0}}V_{t_{1}}\rangle_{AA}, |H_{t_{0}}V_{t_{1}}\rangle_{B^{\prime}B^{\prime}}\}$, which allow the schemes accomplished without post-selection principles.
Moreover, since the time-delay trick allows the undesired terms indistinguishable, the incident photons held by parties keep in the maximally entangled state or in the recyclable less-entangled state.
Though the total success probability of $(2|\alpha\beta|^{2}+\frac{2|\alpha\beta|^{4}}{|\alpha|^{4}+|\beta|^{4}})$ by recycling the less-entangled states is still lower than the nonlinear ECP of Ref. \cite{Sheng2012}, linear optic implementations of our schemes are high-efficient in practice.
However, the inevitable imperfect linear optical elements or dark count will degrade the fidelity of the schemes. Recently, the effect of noise on the measurement is experimentally studied, leading to a reduction in the measurement precision. One effective way to solve this problem is to use the quantum Zeno effect, which can improve the measurement accuracy of entangled probes \cite{QZE}.

In summary, we have presented ECPs for Bell states and GHZ states with unknown parameters.
The schemes are constructed by solely using linear optics, including HWP, PBS,  and single-photon detector.
Our protocols have several characteristics:
First, the protocols can be exactly heralded by the detection signatures, and the photon-number-resolving detections or post-selection principles are not required.
Second, exact parameters $\alpha$ and $\beta$ are not required.
Third, the failed state has a good form for increasing the success probability of the protocols without resorting to the cross-Kerr media.
Fourth, linear optical implementations of the heralded protocols are feasible in the experiment with current technology.
These characteristics make our protocols more useful in long-distance quantum communication.

\medskip

\section*{ACKNOWLEDGEMENTS} \par

This work is supported by  the Fundamental Research Funds for the Central Universities under Grants FRF-TP-19-011A3.



\begin{thebibliography}{0}%
\makeatletter
\providecommand \@ifxundefined [1]{%
 \@ifx{#1\undefined}
}%
\providecommand \@ifnum [1]{%
 \ifnum #1\expandafter \@firstoftwo
 \else \expandafter \@secondoftwo
 \fi
}%
\providecommand \@ifx [1]{%
 \ifx #1\expandafter \@firstoftwo
 \else \expandafter \@secondoftwo
 \fi
}%
\providecommand \natexlab [1]{#1}%
\providecommand \enquote  [1]{``#1''}%
\providecommand \bibnamefont  [1]{#1}%
\providecommand \bibfnamefont [1]{#1}%
\providecommand \citenamefont [1]{#1}%
\providecommand \href@noop [0]{\@secondoftwo}%
\providecommand \href [0]{\begingroup \@sanitize@url \@href}%
\providecommand \@href[1]{\@@startlink{#1}\@@href}%
\providecommand \@@href[1]{\endgroup#1\@@endlink}%
\providecommand \@sanitize@url [0]{\catcode `\\12\catcode `\$12\catcode
  `\&12\catcode `\#12\catcode `\^12\catcode `\_12\catcode `\%12\relax}%
\providecommand \@@startlink[1]{}%
\providecommand \@@endlink[0]{}%
\providecommand \url  [0]{\begingroup\@sanitize@url \@url }%
\providecommand \@url [1]{\endgroup\@href {#1}{\urlprefix }}%
\providecommand \urlprefix  [0]{URL }%
\providecommand \Eprint [0]{\href }%
\providecommand \doibase [0]{https://doi.org/}%
\providecommand \selectlanguage [0]{\@gobble}%
\providecommand \bibinfo  [0]{\@secondoftwo}%
\providecommand \bibfield  [0]{\@secondoftwo}%
\providecommand \translation [1]{[#1]}%
\providecommand \BibitemOpen [0]{}%
\providecommand \bibitemStop [0]{}%
\providecommand \bibitemNoStop [0]{.\EOS\space}%
\providecommand \EOS [0]{\spacefactor3000\relax}%
\providecommand \BibitemShut  [1]{\csname bibitem#1\endcsname}%
\let\auto@bib@innerbib\@empty
\end{thebibliography}%


\medskip


\begin{thebibliography}{99}

\bibitem{1}M. A. Nielsen and I. L. Chuang,  \emph{Quantum Computation and Quantum Information}   (Cambridge University, 2000). 

\bibitem{2}D. P. DiVincenzo, ``Quantum computation,'' Science \textbf{270}(5234), 255-261 (1995). 

\bibitem{3}A. K. Ekert, ``Quantum cryptography based on Bell's theorem,'' Phys. Rev. Lett. \textbf{67}(6), 661-663 (1991).



\bibitem{4}Q. Zhang, F. H. Xu, Y. A. Chen, C. Z. Peng, and J. W. Pan, ``Large scale quantum key distribution: challenges and solutions [Invited],'' Opt. Express \textbf{26}(18), 24260-24273 (2018).

\bibitem{5}S. Olmschenk, D. N. Matsukevich, P. Maunz, D. Hayes, L. M. Duan, and C. Monroe, ``Quantum teleportation between distant matter qubits,'' Science \textbf{323}(5913), 486-489 (2009).

\bibitem{6}S. Pirandola, J. Eisert, C. Weedbrook, A. Furusawa, and S. L. Braunstein, ``Advances in quantum teleportation,'' Nat. Photonics \textbf{9}(10), 641-652 (2015).

\bibitem{7}Y. H. Luo, H. S. Zhong, M. Erhard, X. L. Wang, L. C. Peng, M. Krenn, X. Jiang, L. Li, N. L. Liu, C. Y. Lu, A. Zeilinger, and J. W. Pan, ``Quantum teleportation in high dimensions,'' Phys. Rev. Lett. \textbf{123}(7), 070505 (2019).

\bibitem{8}C. H. Bennett and S. J. Wiesner, ``Communication via one- and two-particle operators on Einstein-Podolsky-Rosen states,'' Phys. Rev. Lett. \textbf{69}(20), 2881-2884 (1992).

\bibitem{9}Y. X. Chen, S. S Liu, Y. B. Lou, and J. T. Jing, ``Orbital angular momentum multiplexed quantum dense coding,'' Phys. Rev. Lett. \textbf{127}(9), 093601 (2021).

\bibitem{10}M. Hillery, V. Bu\v{z}ek, and A. Berthiaume, ``Quantum secret sharing,'' Phys. Rev. A \textbf{59}(3), 1829-1834 (1999).


\bibitem{11}S. M. Lee, S. W. Lee, H. Jeong, and H. S. Park, ``Quantum teleportation of shared quantum secret,'' Phys. Rev. Lett. \textbf{124}(6), 060501 (2020).



\bibitem{12}G. L. Long and X. S. Liu, ``Theoretically efficient highcapacity quantum-key-distribution scheme,'' Phys. Rev. A \textbf{65}(3), 032302 (2002).

\bibitem{13}W. Zhang, D. S. Ding, Y. B. Sheng, L. Zhou, B. S. Shi, and G. C. Guo, ``Quantum secure direct communication with quantum memory,'' Phys. Rev. Lett. \textbf{118}(22), 220501 (2017).

\bibitem{14}Y. B. Sheng, L. Zhou, and G. L. Long, ``One-step quantum secure direct communication,'' Sci. Bull. \textbf{67}(4), 367-374 (2022).

\bibitem{15}C. H. Bennett, G. Brassard, S. Popescu, B. Schumacher, J. A. Smolin, and W. K. Wootters, ``Purification of noisy entanglement and faithful teleportation via noisy channels,'' Phys. Rev. Lett. \textbf{76}(5), 722-725 (1996).

\bibitem{16}C. H. Bennett, H. J. Bernstein, S. Popescu, and B. Schumacher, ``Concentrating partial entanglement by local operations,'' Phys. Rev. A \textbf{53}(4), 2046-2052 (1996).


\bibitem{17}M. Murao, M. B. Plenio, S. Popescu, V. Vedral, and P. L. Knight, ``Multiparticle entanglement purification protocols,'' Phys. Rev. A \textbf{57}(6), R4075-R4078 (1998).

\bibitem{18}Y. B. Sheng, F. G. Deng, and H. Y. Zhou, ``Efficient polarization-entanglement purification based on parametric down-conversion sources with cross-Kerr nonlinearity,'' Phys. Rev. A \textbf{77}(4), 042308 (2008).

\bibitem{19}S. Krastanov, V. V. Albert, and L. Jiang, ``Optimized entanglement purification,'' Quantum \textbf{3}, 123 (2019).

\bibitem{20}X. M. Hu, C. X. Huang, Y. B. Sheng, L. Zhou, B. H. Liu, Y. Guo, C. Zhang, W. B. Xing, Y. F. Huang, C. F. Li, and G. C. Guo, ``Long-distance entanglement purification for quantum communication,'' Phys. Rev. Lett. \textbf{126}(1), 010503 (2021).

\bibitem{21}F. Riera-S\`abat, P. Sekatski, A. Pirker, and W. D\"ur, ``Entanglement-assisted entanglement purification,'' Phys. Rev. Lett. \textbf{127}(4), 040502 (2021).

\bibitem{22}H. X. Yan, Y. P. Zhong, H. S. Chang, A. Bienfait, M. H. Chou, C. R. Conner, \'E. Dumur, J. Grebel, R. G. Povey, and A. N. Cleland, ``Entanglement purification and protection in a superconducting quantum network,'' Phys. Rev. Lett. \textbf{128}(8), 080504 (2022).

\bibitem{23}C. X. Huang, X. M. Hu, B. H. Liu, L. Zhou, Y. B. Sheng, C. F. Li, and G. C. Guo, ``Experimental one-step deterministic polarization entanglement purification,'' Sci. Bull. \textbf{67}(6), 593-597 (2022).


\bibitem{Bose}S. Bose, V. Vedral, and P. L. Knight, ``Purification via entanglement swapping and conserved entanglement,'' Phys. Rev. A \textbf{60}(1), 194-197 (1999).

\bibitem{Shi}B. S. Shi, Y. K. Jiang, and G. C. Guo, ``Optimal entanglement purification via entanglement swapping,'' Phys. Rev. A \textbf{62}(5), 054301 (2000).

\bibitem{Yamamoto}T. Yamamoto, M. Koashi, and N. Imoto, ``Concentration and purification scheme for two partially entangled photon pairs,'' Phys. Rev. A \textbf{64}(1), 012304 (2001).



\bibitem{Zhao}Z. Zhao, J. W. Pan, and M. S. Zhan, ``Practical scheme for entanglement concentration,'' Phys. Rev. A \textbf{64}(1), 014301 (2001).

\bibitem{experiment1}Z. Zhao, T. Yang, Y. A. Chen, A. N. Zhang, and J. W. Pan, ``Experimental realization of entanglement concentration and a quantum repeater,'' Phys. Rev. Lett. \textbf{90}(20), 207901 (2003).

\bibitem{experiment2}T. Yamamoto, M. Koashi, \c{S}. K. \"{O}zdemir, and N. Imoto, ``Experimental extraction of an entangled photon pair from two identically decohered pairs,'' Nature \textbf{421}(6921), 343-346 (2003).


\bibitem{Paunkov}N. Paunkovi\'{c}, Y. Omar, S. Bose, and V. Vedral, ``Entanglement concentration using quantum statistics,'' Phys. Rev. Lett. \textbf{88}(18), 187903 (2002).

\bibitem{Sheng2008}Y. B. Sheng, F. G. Deng, and H. Y. Zhou, ``Nonlocal entanglement concentration scheme for partially entangled multipartite systems with nonlinear optics,'' Phys. Rev. A \textbf{77}(6), 062325 (2008).

\bibitem{Sheng2012}Y. B. Sheng, L. Zhou, S. M. Zhao, and B. Y. Zheng, ``Efficient single-photon-assisted entanglement concentration for partially entangled photon pairs,'' Phys. Rev. A \textbf{85}(1), 012307 (2012).

\bibitem{Wang2010}H. F. Wang, S. Zhang, and K. H. Yeon, ``Linear optical scheme for entanglement concentration of twopartially entangled three-photon ${ W}$ states,'' Eur. Phys. J. D \textbf{56}(2), 271-275 (2010).

\bibitem{Yildiz}A. Yildiz, ``Optimal distillation of three-qubit $W$ states,'' Phys. Rev. A \textbf{82}(1), 012317 (2010).

\bibitem{Sheng2012w}Y. B. Sheng, L. Zhou, and S. M. Zhao, ``Efficient two-step entanglement concentration for arbitrary $W$ states,'' Phys. Rev. A \textbf{85}(4), 042302 (2012).

\bibitem{Yan2014}X. Yan, Y. F. Yu, and Z. M. Zhang, ``Entanglement concentration for a non-maximally entangled four-photon cluster state,'' Front. Phys. \textbf{9}(5), 640–645 (2014).


\bibitem{Sheng2015w}Y. B. Sheng, J. Pan, R. Guo, L. Zhou, and L. Wang, ``Efficient $N$-particle $W$ state concentration with different parity check gates,'' Sci. China Phys. Mech. Astron. \textbf{58}(6), 060301 (2015).


\bibitem{Zhang2017}H. Zhang and H. B. Wang, ``Entanglement concentration of microwave photons based on the Kerr effect in circuit QED,'' Phys. Rev. A \textbf{95}(5), 052314 (2017).

\bibitem{split}B. C. Ren, F. F. Du, and F. G. Deng, ``Hyperentanglement concentration for two-photon four-qubit systems with linear optics,'' Phys. Rev. A \textbf{88}(1), 012302 (2013).

\bibitem{hyper00}X. H. Li, X. Chen, and Z. Zeng,  ``Hyperconcentration for entanglement in two degrees of freedom,'' J. Opt. Soc. Am. B \textbf{30}(11), 2774–2780 (2013).

\bibitem{hyper01}X. Chen, Z. Zeng, and X. H. Li, ``Hyperconcentration based on projection measurements,'' Commun. Theor. Phys. \textbf{61}(3), 322-328 (2014).

\bibitem{hyper1}X. H. Li and S. Ghose, ``Hyperentanglement concentration for time-bin and polarization hyperentangled photons,'' Phys. Rev. A \textbf{91}(6), 062302 (2015).

\bibitem{hyper2}X. H. Li and S. Ghose, ``Efficient hyperconcentration of nonlocal multipartite entanglement via the cross-Kerr nonlinearity,'' Opt. Express \textbf{23}(3), 3550-3562 (2015).

\bibitem{hyper3}H. Wang, B. C. Ren, A. H. Wang, A. Alsaedi, T. Hayat, and F. G. Deng, ``General hyperentanglement concentration for polarization-spatial-time-bin multi-photon systems with linear optics,'' Front. Phys. \textbf{13}(5), 130315 (2018).

\bibitem{hyper4}C. Y. Li and Y. Shen, ``Asymmetrical hyperentanglement concentration for entanglement of polarization and orbital angular momentum,'' Opt. Express \textbf{27}(9), 13172-13181 (2019).

\bibitem{QAZE1}Q. Ai, Y. Li, H. Zheng, and C. P. Sun, ``Quantum anti-Zeno effect without rotating wave approximation,'' Phys. Rev. A \textbf{81}(4), 042116 (2010).

\bibitem{QAZE2}D. Z. Xu, Q. Ai, and C. P. Sun, ``Dispersive-coupling-based quantum Zeno effect in a cavity-QED system,'' Phys. Rev. A \textbf{83}(2), 022107 (2011).

\bibitem{QZE}X. Y. Long, W. T. He, N. N. Zhang, K. Tang, Z. D. Lin, H. F. Liu, X. F. Nie, G. R. Feng, J. Li, T. Xin,
Q. Ai, and D. W. Lu, ``Entanglement-Enhanced Quantum Metrology in Colored Noise by Quantum Zeno Effect,'' Phys. Rev. Lett. \textbf{129}(7), 070502 (2021).


\end{thebibliography}
\end{document}